%% file: main_camera_ready.tex
\documentclass[10pt,twocolumn,letterpaper]{article}

\usepackage{cvpr}              


\definecolor{cvprblue}{rgb}{0.21,0.49,0.74}
\usepackage[breaklinks,colorlinks,allcolors=cvprblue]{hyperref}

\usepackage{amsmath,amssymb,mathtools,bm}
\usepackage{graphicx}
\usepackage{booktabs}
\usepackage{multirow}
\usepackage{microtype}
\usepackage{siunitx}
\usepackage{enumitem}
\usepackage{xcolor}
\usepackage{url}
\usepackage{placeins}
\usepackage{tabularx}
\usepackage{float}

\definecolor{posgreen}{RGB}{0,130,60}
\definecolor{negred}{RGB}{200,30,30}

\def\confName{CVPR}
\def\confYear{2026}

\title{When Prompts Mislead: Textual Dominance and Diagnostic Bias in MLLMs\vspace{-0.6em}}

\author{
Inhyuk Park \quad Doohyun Park\thanks{Corresponding author}\\[-1pt]
VUNO Inc.\\[-2pt]
{\tt\small \{inhyuk.park, doohyun.park\}@vuno.co}
}

\begin{document}

\maketitle

\input{sec/0_abstract}
\input{sec/1_intro}
\input{sec/2_relatedwork}
\input{sec/3_methods}
\input{sec/4_results}
\input{sec/5_limit_future}
\input{sec/6_conclusion}

{
    \small
    \bibliographystyle{ieeenat_fullname}
    \bibliography{main}
}


\end{document}

%% file: sec/0_abstract.tex
\begin{abstract}
Multimodal large language models (MLLMs) are increasingly being evaluated for medical applications, where computational constraints often make prompting strategies the only practical alternative to fine-tuning. Such strategies are generally assumed to support diagnostic reasoning, yet their potential failure modes in medical MLLMs remain poorly characterized. We analyze FundusExpert-1B, an open-source ophthalmology MLLM, on a hemorrhage versus drusen discrimination task using the public BRSET dataset, adopted here as a controlled testbed for our analysis. (i) A controlled probe with artificially injected markers confirms that the model retains coarse, region-level spatial grounding. (ii) Compared with zero-shot inference, one-shot textual prompts bias predictions toward the prompted finding. (iii) When an overlaid lesion contour is paired with an inconsistent textual claim, the textual prompt overrides the correct visual cue: overall accuracy drops from 75\% to 46\% relative to the visual-only condition, and Chain-of-Thought (CoT) reasoning is associated with further degradation rather than self-correction. Although limited to a single model and dataset, our findings suggest that prompting strategies alone may be insufficient for the safe clinical deployment of medical MLLMs.
\end{abstract}

%% file: sec/1_intro.tex
\section{Introduction}
\label{sec:intro}

\begin{figure*}[t]
\centering
\includegraphics[width=\textwidth]{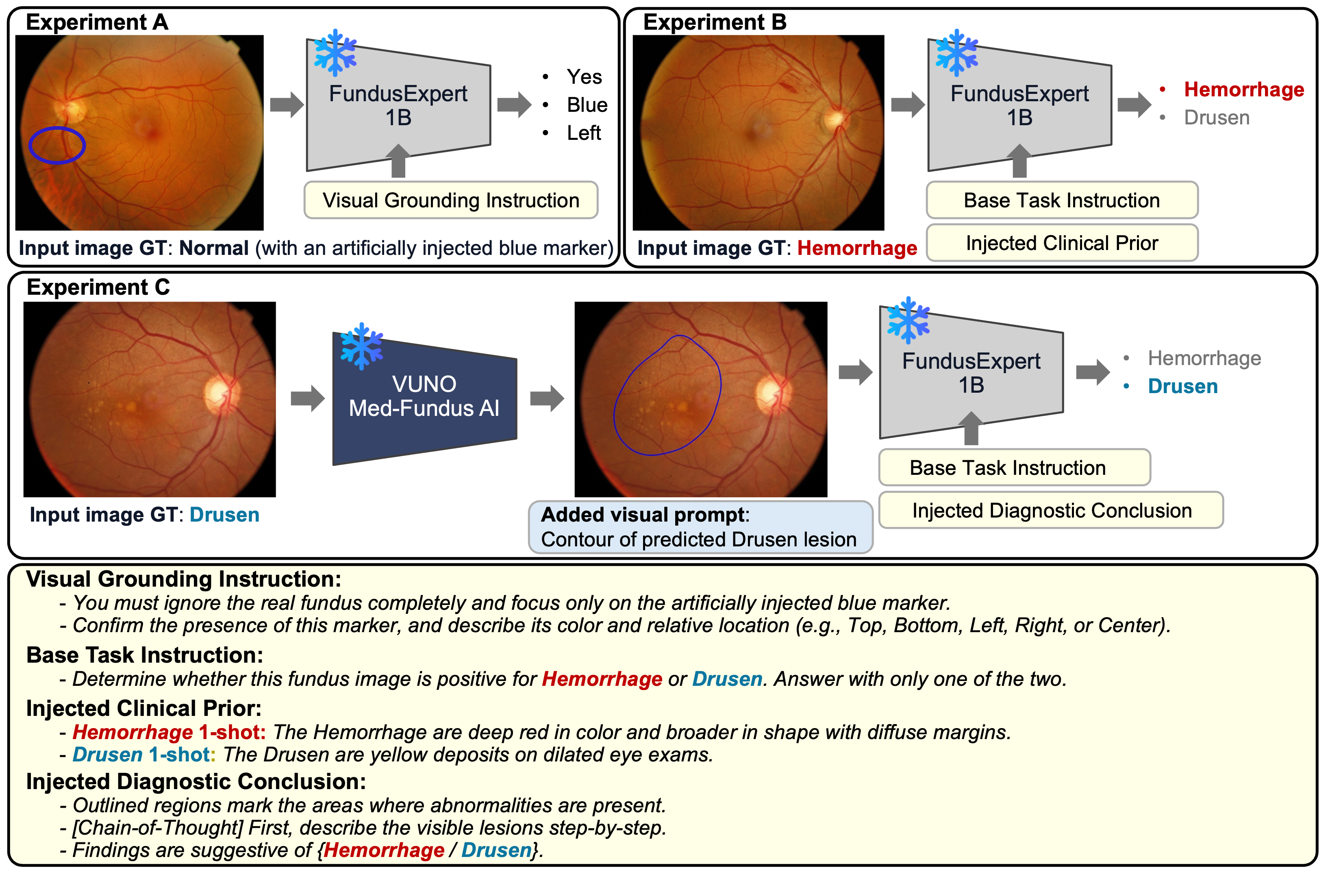}
\caption{\textbf{Three-stage evaluation pipeline on a frozen FundusExpert-1B.} \textbf{(A)} Visual grounding probe: an artificially injected blue marker is overlaid on a normal fundus image, and the model is queried for the marker's presence, color, and approximate location. \textbf{(B)} Diagnostic discrimination (\textcolor{red}{Hemorrhage} vs.\ \textcolor{blue}{Drusen}) with a one-shot clinical description supplied as a textual prior. \textbf{(C)} Multimodal prompting: an overlaid lesion contour from a commercial fundus AI is provided as a visual prompt together with an assertive diagnostic claim as a textual prompt, optionally with a Chain-of-Thought (CoT) instruction.}
\label{fig:pipeline}
\end{figure*}

Multimodal large language models (MLLMs) have shown strong potential in medical image analysis, with a growing body of work specializing them to clinical domains~\cite{nath2025vila,chen2025mimo,liu2025constructing,chen2025deep}. In practice, however, clinical deployment is typically constrained by limited computational resources, which makes the operation of large proprietary models or full-parameter fine-tuning impractical~\cite{khan2025comprehensive}. As a result, inference with frozen parameters combined with prompting strategies has emerged as a common adaptation approach~\cite{liu2023pre,taylor2023clinical}.

Two complementary forms of prompting are common in this setting. \emph{Visual} prompts, such as bounding boxes or contours, direct the model toward regions of interest~\cite{denner2024visual,wu2024one,chen2025mimo}. \emph{Textual} prompts, such as one-shot exemplars or clinical guidelines, inject task-relevant priors~\cite{bu2024instance}. Both are generally assumed to be additive: the model is expected to integrate the prompt with the visual evidence and produce a more reliable diagnosis.

Recent work outside the medical domain suggests that this assumption does not always hold. MLLMs frequently exhibit object hallucinations~\cite{leng2024mitigating,jiang2024hallucination} and visual shortcomings~\cite{tong2024eyes}, and their predictions can be biased by language priors over visual evidence~\cite{zhang2025debiasing}. Whether these failure modes also arise in medical MLLMs adapted through prompting alone, where the cost of incorrect diagnoses is high, has not been systematically examined.

In this work, we characterize prompt-induced failure modes in FundusExpert~\cite{liu2025constructing}, an open-source ophthalmology MLLM that reports strong zero-shot performance over GPT-4o~\cite{hurst2024gpt} (77.0\% vs.\ 47.6\% clinical agreement at 8B). FundusExpert currently represents the frontier of open-source ophthalmology MLLMs, outperforming existing models across multiple clinical benchmarks. We focus on the lightweight 1B variant, which is representative of resource-constrained edge environments and still substantially outperforms GPT-4o on Fundus-MMBench~\cite{liu2025constructing} (63.5\% vs.\ 41.6\%). We restrict the evaluation to hemorrhage and drusen, two findings explicitly and abundantly represented in both BRSET~\cite{nakayama2023brazilian} and the model's training data (FundusGen~\cite{liu2025constructing}). This setup provides a controlled testbed for isolating prompt-induced effects. Our analysis proceeds in three stages:

\begin{enumerate}[label=\arabic*., leftmargin=*]
    \item A controlled probe with artificially injected markers confirms that the model retains the coarse, region-level spatial grounding required to localize and respond to visually marked findings.
    \item Compared with zero-shot inference, one-shot textual prompts bias predictions toward the prompted finding.
    \item When textual prompts contradict explicit visual cues, the model follows the text, and Chain-of-Thought (CoT) reasoning is associated with further degradation rather than self-correction.
\end{enumerate}

Overall, this work provides the focused characterization of prompt-induced failure modes, particularly textual dominance under explicit modality conflict, in medical MLLMs adapted purely through prompting strategies.

%% file: sec/2_relatedwork.tex
\section{Related Work}
\label{sec:related}

\subsection{Medical MLLMs and Domain Adaptation}
Domain-specific MLLMs have been actively developed for medical imaging. VILA-M3~\cite{nath2025vila} augments general VLMs with medical expert knowledge. MIMO~\cite{chen2025mimo} integrates visual referring inputs (e.g., region masks) with pixel-grounded outputs for medical visual question answering. In ophthalmology, FundusExpert~\cite{liu2025constructing} introduces a clinical cognitive chain combining region-based feature analysis with diagnostic reasoning. These efforts have largely been benchmarked under standard zero-shot or instruction-tuned settings, where prompt design is treated as a means to elicit performance rather than as a source of systematic bias. While these works demonstrate the feasibility of medical MLLMs, they primarily focus on architectural design and training strategies, leaving the behavior of frozen medical MLLMs under prompt-based adaptation underexplored.

\subsection{Prompt-based Adaptation in Medical Imaging}
Various strategies have been proposed to improve medical foundation models through prompting rather than retraining. On the visual side, bounding boxes and arrows can focus zero-shot models on local lesions~\cite{denner2024visual}, and a one-prompt paradigm has been shown to generalize across medical segmentation tasks~\cite{wu2024one}. On the textual side, DKP~\cite{bu2024dynamic} dynamically extracts lesion-level information and injects it as textual guidelines. These works generally assume that prompts have an additive effect, with limited attention to failure cases when prompts are insufficient or misleading.

\subsection{Failure Modes of MLLMs}
A separate line of research has identified failure modes in general-domain MLLMs: systematic visual shortcomings where the model fails to perceive details evident in the image~\cite{tong2024eyes}; hallucination, addressed through contrastive decoding~\cite{leng2024mitigating} and augmented contrastive training~\cite{jiang2024hallucination}; and modality bias, over-reliance on either textual priors or visual evidence, in the natural-image domain~\cite{zhang2025debiasing}. Whether and how such textual dominance arises in medical MLLMs adapted purely through prompting strategies has not been directly examined, particularly under explicit modality conflict; the present work addresses this gap.

%% file: sec/3_methods.tex
\section{Methods}
\label{sec:methods}
We analyze prompt-induced failure modes in FundusExpert-1B~\cite{liu2025constructing} through three controlled experiments. The model is used in inference mode with all parameters kept frozen and with deterministic decoding.

\subsection{Dataset and Task Formulation}
\label{sec:methods:dataset}
We use BRSET~\cite{nakayama2023brazilian}, a publicly available fundus dataset, focusing on hemorrhage and drusen. Cases positive for both findings (co-positive) are excluded to avoid label ambiguity. For each finding, we construct a balanced 100-image evaluation set: 50 target-positive images and 50 controls positive for the alternative finding only. The hemorrhage and drusen sets are sampled independently. All diagnostic queries use a fixed forced-choice prompt: \textit{Determine whether this fundus image is positive for Hemorrhage or Drusen. Answer with only one of the two.}

\subsection{Probing Fundamental Visual Grounding}
\label{sec:methods:probe}
To verify that the model can in principle localize and respond to visually marked findings, we run a controlled probe on 30 normal BRSET images without pathology. Each image is overlaid with an artificially injected blue marker matching the contour color used in Sec.~\ref{sec:methods:multimodal}. The model is queried with a strict-rule prompt that forbids any retinal description (e.g., \textit{ignore the real fundus completely}) and asks only about the marker. Each image is queried independently for the marker's existence, color, and approximate location among five coarse regions (Top, Bottom, Left, Right, Center), yielding 90 evaluations across 30 images.

\subsection{Prompt-Induced Bias}
\label{sec:methods:bias}
To test whether textual prompts bias predictions, we compare zero-shot inference to one-shot prompting on each task-specific set. We construct two one-shot prompts adapted from clinical reference texts: for hemorrhage, \textit{The Hemorrhage are deep red in color and broader in shape with diffuse margins}~\cite{Kanukollu2026RetinalHemorrhage}; for drusen, \textit{The Drusen are yellow deposits on dilated eye exams}~\cite{VanDenLangenberg2026DrusenBodies}. Each prompt is applied to both sets, allowing us to measure the effect on both the prompted (target) finding and the alternative (off-target) finding.

\subsection{Multimodal Prompting and Textual Dominance}
\label{sec:methods:multimodal}
To test how the model resolves modality conflict, we combine visual prompts (lesion contours) with short textual statements of the form \textit{Findings are suggestive of [Hemorrhage/Drusen]}. Lesion contours are extracted by a commercial fundus analysis system (VUNO Med-Fundus AI), which provides automated detection for both findings; for each image we use the system's predicted contour for the ground-truth finding, so that the contour, by construction, localizes the actual finding. Contour fidelity was not separately validated against expert annotation; we revisit this point in Sec.~\ref{sec:limitations}. Unlike the indirect clinical descriptions in Sec.~\ref{sec:methods:bias}, the textual prompt here delivers an assertive diagnostic claim, thereby maximizing textual-visual conflict in the inconsistent condition. We define five conditions:
\begin{itemize}
    \item \textbf{Baseline:} the original image, without prompts.
    \item \textbf{Visual-only:} image with the lesion contour overlay.
    \item \textbf{Consistent:} contour plus a textual statement matching the visual cue, serving as an upper bound under aligned modalities.
    \item \textbf{Inconsistent:} contour plus a textual statement contradicting the visual cue, used to quantify \emph{textual dominance}.
    \item \textbf{CoT~\cite{wei2022chain}:} the Inconsistent condition prepended with \textit{First, describe the visible lesions step-by-step}, testing whether stepwise reasoning enables self-correction under a misleading textual prompt.
\end{itemize}
We operationalize \emph{textual dominance} as the drop in overall accuracy between the Visual-only and Inconsistent conditions: a positive drop indicates that the inconsistent textual statement overrides the correct visual evidence localized by the contour.

%% file: sec/4_results.tex
\section{Results}
\label{sec:results}

\noindent\textbf{Visual grounding} (Tab.~\ref{tab:visual_grounding}). The model correctly detects the artificially injected marker and identifies its color in all 30 images, and localizes it correctly in 27 of 30 (90\%). This confirms that the model retains the coarse, region-level spatial grounding required to localize and respond to visually marked findings; the failures observed below are therefore unlikely to reflect a fundamental absence of spatial grounding.

\begin{table}[H]
\centering
\caption{Quantitative results of fundamental visual grounding on 30 normal fundus images.}
\label{tab:visual_grounding}
\small
\begin{tabular*}{0.96\columnwidth}{@{\extracolsep{\fill}}llc@{}}
\toprule
\textbf{Category} & \textbf{Metric} & \textbf{Correct Images} \\
\midrule
Presence & Detection & 30 \\
\midrule
Attribute & Color & 30 \\
\midrule
Spatial & Location & 27 \\
\bottomrule
\end{tabular*}
\end{table}

\noindent\textbf{Prompt-induced bias} (Tab.~\ref{tab:oneshot-bias}). The one-shot prompts contain only neutral clinical descriptions of each finding and convey no diagnostic suggestion about the input image. Nevertheless, predictions are systematically biased toward the described finding in both tasks. When the description matches the target finding, sensitivity rises (from 0.82 to 0.90 for drusen and from 0.34 to 0.36 for hemorrhage); when it matches the alternative finding, sensitivity falls (from 0.82 to 0.72 for drusen and from 0.34 to 0.22 for hemorrhage). These two patterns reflect a single underlying mechanism rather than two distinct effects: in both cases, the model shifts its output toward the described finding regardless of the actual visual evidence.

\begin{table}[H]
\centering
\caption{Prompt-induced bias in frozen fundus MLLM inference. Each task consists of 50 target-positive and 50 alternative-positive samples. Sens., Spec., and Bal-Acc denote sensitivity, specificity, and balanced accuracy, respectively. Values in parentheses indicate the change relative to the zero-shot baseline.}
\label{tab:oneshot-bias}
\small
\setlength{\tabcolsep}{4pt}
\begin{tabular}{ll r@{\,}l r@{\,}l r@{\,}l}
\toprule
\textbf{Task} & \textbf{Prompt} & \multicolumn{2}{c}{\textbf{Sens.}} & \multicolumn{2}{c}{\textbf{Spec.}} & \multicolumn{2}{c}{\textbf{Bal-Acc}} \\
\midrule
Hemorrhage 
& Zero-shot     & 0.34 &                                              & 0.58 &                                              & 0.46 & \\
& Hemor. 1-shot & 0.36 & \tiny{\textcolor{posgreen}{(+0.02)}} & 0.66 & \tiny{\textcolor{posgreen}{(+0.08)}} & 0.51 & \tiny{\textcolor{posgreen}{(+0.05)}} \\
& Drusen 1-shot & 0.22 & \tiny{\textcolor{negred}{(-0.12)}}    & 0.60 & \tiny{\textcolor{posgreen}{(+0.02)}} & 0.41 & \tiny{\textcolor{negred}{(-0.05)}} \\
\addlinespace
\midrule
Drusen
& Zero-shot     & 0.82 &                                              & 0.90 &                                              & 0.86 & \\
& Drusen 1-shot & 0.90 & \tiny{\textcolor{posgreen}{(+0.08)}} & 0.94 & \tiny{\textcolor{posgreen}{(+0.04)}} & 0.92 & \tiny{\textcolor{posgreen}{(+0.06)}} \\
& Hemor. 1-shot & 0.72 & \tiny{\textcolor{negred}{(-0.10)}}    & 0.94 & \tiny{\textcolor{posgreen}{(+0.04)}} & 0.83 & \tiny{\textcolor{negred}{(-0.03)}} \\
\bottomrule
\end{tabular}
\end{table}

\noindent\textbf{Textual dominance} (Tab.~\ref{tab:cognitive_collapse}). Adding the lesion contour alone (Visual-only) raises overall accuracy from 0.58 (Baseline) to 0.75, confirming that the model can use the contour as visual guidance. Adding a Consistent textual prompt further raises accuracy to 0.85. An Inconsistent textual prompt instead drops accuracy to 0.46, an absolute drop of 0.29 from the Visual-only condition, even though the contour, by construction, localizes the actual finding. We take this 0.29 drop as the empirical measure of textual dominance (Sec.~\ref{sec:methods:multimodal}). Adding CoT to the Inconsistent condition further reduces overall accuracy to 0.36, with the hemorrhage-specific sensitivity collapsing to 0.08, representing only 4 of 50 hemorrhage cases correctly classified despite the contour overlay. Stepwise reasoning therefore fails to recover the visual evidence; the pattern is consistent with the model verbalizing a description that aligns with the false textual claim rather than self-correcting.

\begin{table}[H]
\centering
\caption{Ablation study of visual and textual prompting. Hem.\ and Drusen columns report per-class sensitivity (50 cases per class); Overall reports overall accuracy across both classes (100 cases). Values in parentheses indicate the change relative to the Visual-only condition.}
\label{tab:cognitive_collapse}
\small
\setlength{\tabcolsep}{4pt}
\begin{tabular}{cccc | r@{\,}l r@{\,}l r@{\,}l}
\toprule
\textbf{Image} & \textbf{Visual} & \textbf{Text} & \textbf{CoT} & \multicolumn{2}{c}{\textbf{Hem.}} & \multicolumn{2}{c}{\textbf{Drusen}} & \multicolumn{2}{c}{\textbf{Overall}} \\
\midrule
\checkmark &            &        &            & 0.34 &                                                  & 0.82 &                                                  & 0.58 & \\
\checkmark & \checkmark &        &            & 0.56 &                                                  & 0.94 &                                                  & 0.75 & \\
\checkmark & \checkmark & Cons.  &            & 0.72 & \tiny{\textcolor{green!50!black}{(+0.16)}} & 0.98 & \tiny{\textcolor{green!50!black}{(+0.04)}} & 0.85 & \tiny{\textcolor{green!50!black}{(+0.10)}} \\
\checkmark & \checkmark & Incons.&            & 0.26 & \tiny{\textcolor{red!80!black}{(-0.30)}}    & 0.66 & \tiny{\textcolor{red!80!black}{(-0.28)}}    & 0.46 & \tiny{\textcolor{red!80!black}{(-0.29)}} \\
\checkmark & \checkmark & Incons.& \checkmark & 0.08 & \tiny{\textcolor{red!80!black}{(-0.48)}}    & 0.64 & \tiny{\textcolor{red!80!black}{(-0.30)}}    & 0.36 & \tiny{\textcolor{red!80!black}{(-0.39)}} \\
\bottomrule
\end{tabular}
\end{table}

%% file: sec/5_limit_future.tex
\section{Limitations and Future Work}
\label{sec:limitations}

\noindent \textbf{Scope of model and dataset.} We evaluate a single model (FundusExpert-1B) on a single dataset (BRSET) and on two retinal findings. The near-chance hemorrhage zero-shot baseline, the asymmetry between findings, and the magnitude of textual dominance may differ in larger medical MLLMs (e.g., Med-Gemini~\cite{saab2024capabilities}, LLaVA-NeXT-Med~\cite{guo2025llava}, MedGemma~\cite{sellergren2025medgemma}), in other fundus datasets, or on findings less abundantly represented in training. The present results should therefore be read as a case study rather than general claims about medical MLLMs. Nevertheless, characterizing these failure modes in a model that currently defines the performance frontier highlights reliability risks inherent in even high-performing specialized MLLMs.

\noindent \textbf{External dependencies and probe design.} The visual prompt depends on contours generated by VUNO Med-Fundus AI, whose detection fidelity was not separately validated against expert annotation; lower-quality contours would weaken the contradiction in the Inconsistent condition and could either inflate or attenuate the observed dominance. The visual grounding probe (Sec.~\ref{sec:methods:probe}) further evaluates only coarse, region-level localization on an artificially injected marker and lacks marker-absent negative controls, so it does not directly measure fine-grained region integration or prompt-conditioned hallucination.

\noindent \textbf{Future work.} We plan to (i) extend the analysis across multiple ophthalmology MLLMs of varying scale and across additional fundus datasets and findings; (ii) validate contour quality with expert annotation and incorporate marker-absent negative controls in the grounding probe; and (iii) investigate lightweight alignment or guardrail techniques that explicitly mitigate textual dominance under modality conflict.

%% file: sec/6_conclusion.tex
\noindent \section{Conclusion}
We examined how visual and textual prompts shape predictions in a frozen lightweight medical MLLM. The model retains the coarse spatial grounding needed to act on cued visual evidence, yet its predictions are markedly responsive to textual context. Even neutral one-shot descriptions bias predictions toward the described finding in both tasks, indicating that the model treats textual cues as strong priors rather than as supplementary clinical context. When textual and visual cues conflict, the textual signal overrides the correct visual evidence, and Chain-of-Thought reasoning is associated with further degradation rather than self-correction. Although based on a single model and dataset and best regarded as a case study, these results suggest that prompting strategies alone may be insufficient for the safe clinical deployment of frozen medical MLLMs; more fundamental interventions, such as lightweight alignment or guardrails that constrain modality conflict and prompt-conditioned bias, may be necessary.